\documentclass[aps,prl,twocolumn,superscriptaddress,showpacks]{revtex4-2}

\usepackage{comment}
\usepackage[colorlinks, citecolor=red]{hyperref}
\usepackage{hyperref}
\usepackage[utf8]{inputenc}
\usepackage[T1]{fontenc}    %
\usepackage[english]{babel}
\usepackage[pdftex]{graphicx}
\usepackage{amsmath,amssymb,amsthm,bbm, mathtools} %
\usepackage{mathrsfs}
\usepackage{tikz}   %
\usepackage[normalem]{ulem}

\renewcommand{\vec}[1]{\boldsymbol{#1}}

\begin{document}

%\title{Zero-temperature resistance from decoherence in bulk topological insulators}

%\title{Topological metal}
\title{Metalization of topological insulators}

%\title{Quantum geometric metal}

\author{Xian-Peng Zhang}
\affiliation{Centre for Quantum Physics, Key Laboratory of Advanced Optoelectronic Quantum Architecture and Measurement (MOE), School of Physics, Beijing Institute of Technology, Beijing, 100081, China}

\affiliation{International Center for Quantum Materials, Beijing Institute of Technology, Zhuhai, 519000, China}

\author{Yan-Qing Feng}
%\email{yq_feng@bitzh.edu.cn}

\affiliation{International Center for Quantum Materials, Beijing Institute of Technology, Zhuhai, 519000, China}

\author{Ji-Feng Shao}
%\email{shaojifeng@bitzh.edu.cn}

\affiliation{International Center for Quantum Materials, Beijing Institute of Technology, Zhuhai, 519000, China}

\author{Haiwen Liu}
\affiliation{Center for Advanced Quantum Studies, School of Physics and Astronomy, Beijing Normal University, Beijing 100875, China}

\author{Yugui Yao}
\email{ygyao@bit.edu.cn}
\affiliation{Centre for Quantum Physics, Key Laboratory of Advanced Optoelectronic Quantum Architecture and Measurement (MOE), School of Physics, Beijing Institute of Technology, Beijing, 100081, China}

\affiliation{International Center for Quantum Materials, Beijing Institute of Technology, Zhuhai, 519000, China}

\begin{abstract} 
In modern condensed matter theory, phases of electronic matter--such as metals and insulators--are fundamentally distinguished by the presence or absence of charge-carrying quasiparticles or excitations near the Fermi surface at low temperatures. Here, we show that this criterion breaks down in Berry-curvature-dominated systems, where transport is governed by interband coherence across the entire Fermi sea. We develop a microscopic theory of quantum transport in bulk topological insulators with a vanishing density of states at the Fermi energy, for which the conventional Drude contribution is absent. We demonstrate that impurity-scattering–induced coherence decay generates a distinct longitudinal transport channel even in the topologically trivial regime, with edge contributions rigorously excluded. This mechanism yields a finite longitudinal conductivity even in the absence of carriers at the Fermi level and exhibits an unconventional scaling linear in impurity density in the dilute limit, in stark contrast to Drude behaviour. Importantly, this decoherence-induced  conductance is inversely proportional to temperature, reminiscent of strange-metal behaviour, most prominently observed in cuprate superconductors above their critical temperature. Our findings reveal quantum decoherence as a fundamental origin of longitudinal transport beyond the Drude paradigm, challenging the traditional distinction between metals and insulators.
\end{abstract}

\maketitle

\textit{Introduction--}The standard distinction between metals and insulators is rooted in band theory~\cite{kittel2018introduction,mahan2013many,landau2013course,peierls1955quantum,scalapino1993insulator}, where electrical conduction originates from Fermi-surface quasiparticles/excitations undergoing momentum relaxation—the decay of the diagonal components of the density matrix. Metals host gapless charge excitations and therefore exhibit finite conductivity in the zero-temperature limit. Insulators, by contrast, possess a bulk energy gap between a filled valence band and an empty conduction band; lacking such excitations, their conductivity vanishes as temperature approaches zero. This picture is first challenged by Mott insulators: although band theory predicts metallic behaviour, strong electron–electron interactions localize charge carriers and open a correlation-driven gap~\cite{mott1937discussion,mott1949basis,hubbard1963electron}. More broadly, this bulk-based classification fails to capture boundary-dominated transport. In topological systems, symmetry-protected edge or surface states can provide robust conduction channels even when the bulk is fully gapped~\cite{qi2011topological,hasan2010colloquium,moore2010birth}. Consequently, finite longitudinal conductivity persists at low temperatures despite the absence of bulk carriers~\cite{konig2007quantum,knez2010finite,shamim2021quantized}.

\begin{figure}[t!]
\begin{center}
\includegraphics[width=0.48\textwidth]{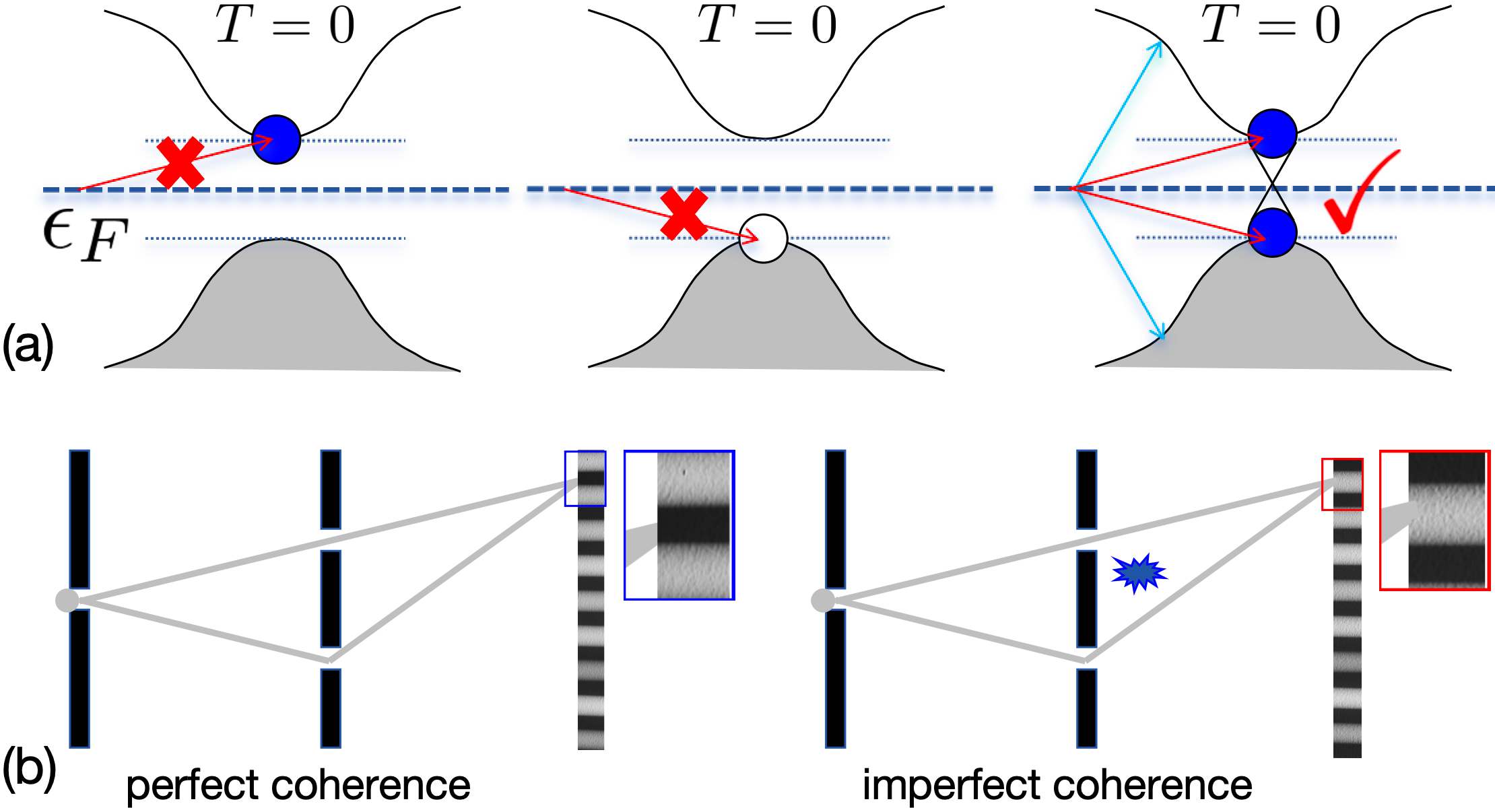} 
\end{center}
\caption{(a) Coherent transport pathways: when an electron occupies coherent superpositions of conduction- and valence-band states, it can effectively bridge the large bulk gap of the topological insulator. (b) The double-slit interference experiment: under perfectly coherent conditions, destructive interference initially produces a dark fringe, whereas introducing disturbances can lead to the appearance of bright fringe.}
\label{FIGS}
\end{figure}

Our recent work implies a more fundamental breakdown of this paradigm  in Berry-curvature-dominated systems~\cite{zhang2026magnetoresistance,zhang2026theory,zhang2026theoryti}, where an electric field induces appreciable interband coherence throughout the entire Fermi sea~\cite{culcer2017interband,sekine2017quantum,atencia2022semiclassical}. In such systems, transport cannot be described solely in terms of Fermi-surface quasiparticles. Instead, quantum coherence between bands--encoded in the off-diagonal elements of the density matrix--emerges as an additional dynamical degree of freedom~\cite{zhang2026magnetoresistance,zhang2026theory,zhang2026theoryti,culcer2017interband,sekine2017quantum,atencia2022semiclassical}.  Crucially, the decay of this coherence does not simply suppress coherent effects~\cite{zhang2026theoryti,zhang2026theory}; rather, it gives rise to a distinct longitudinal transport channel~\cite{zhang2026magnetoresistance}. In sharp contrast to the Drude conductivity inversely proportional to impurity density, the decoherence-induced contribution scales linearly with impurity density in the dilute  limit~\cite{zhang2026magnetoresistance}. This seemingly paradoxical scaling is, in fact, allowed, consistent with early studies of quantum linear magnetoresistance in strong orbital magnetic fields by Abrikosov~\cite{abrikosov1969galvanomagnetic,abrikosov1998quantum,abrikosov2003quantum} and Bastin~\cite{bastin1971quantum}. Consequently, the Drude and decoherence mechanisms therefore exhibit parametrically opposite disorder dependences:  $\sigma_{\text{dia}}\propto 1/n_i$~\cite{drude1900elektronentheorie} while $\sigma_{\text{off}}\propto n_{\mathrm{i}}$~\cite{zhang2026magnetoresistance}. Their physical origins are inherently distinct. Drude transport is a Fermi-surface property, insensitive to the quantum geometry of Bloch states. By contrast, the decoherence-driven contribution is a Fermi-sea effect controlled by Berry connection and interband matrix elements. In Berry-curvature-dominated systems~\cite{xiao2010berry}, the decay of these coherences converts geometric properties of the wavefunctions into a dissipative longitudinal response~\cite{zhang2026magnetoresistance}. Remarkably, this mechanism persists even when the Fermi energy lies within a bulk gap, thereby invalidating the conventional association between a vanishing density of states and insulating behaviour. The total conductivity therefore separates into two competing channels $\sigma=a\frac{\nu_F}{n_i}+bn_i$~\cite{zhang2026magnetoresistance,zhang2026theory,zhang2026theoryti}, which have parametrically distinct disorder dependences in the dilute impurity limit. This decomposition signals a breakdown of the conventional clean-band criterion for insulating behaviour and predicts disorder-enhanced conductivity in regimes where semiclassical quasiparticle transport is suppressed. Experimentally, however, isolating the decoherence-induced contribution remains challenging, as both channels coexist in realistic samples.

To expose the decoherence channel, we consider macroscopic topological insulators with vanishing bulk density of states at the Fermi level, where the Drude contribution is suppressed. Within a quantum master equation framework, we develop a microscopic theory of decoherence-induced longitudinal resistance. Focusing on HgTe/CdTe quantum wells, we identify two distinct mechanisms by which coherence decay generates finite resistance and propose clear experimental signatures—most notably, a linear scaling with impurity density in the dilute limit, in stark contrast to the Drude contribution. This quantum resistance persists even in the topologically trivial regime, with edge contributions rigorously excluded, providing a direct signature of decoherence-driven transport. Our results uncover a Berry-curvature-driven origin of longitudinal response beyond the Drude paradigm and call for a re-examination of the conventional classification of metals, semiconductors, and insulators.

\textit{Model and theory--} We work on the nonequilibrium quantum transport of itinerant electrons governed by the total Hamiltonian $\hat{H}=\hat{H}_{e}+\hat{H}_{E}+\hat{V}$~\cite{zhang2026theoryti}. The electronic structure is described by the Bernevig–Hughes–Zhang model,   $\hat{H}_{e}=sAk_x\hat{\sigma}_x+Ak_y\hat{\sigma}_y+(M-Bk^2)\hat{\sigma}_z$~\cite{qi2010quantum,bernevig2006quantumexp}, which realises a two-dimensional topological insulator for $MB>0$~\cite{konig2007quantum,nowack2013imaging}. Here, $\vec{\sigma}=(\hat{\sigma}_x,\hat{\sigma}_y,\hat{\sigma}_z)$denote Pauli matrices in orbital space. The coupling to an external electric field is given by $\hat{H}_{E}=- e\vec{E}\cdot \hat{\vec{r}}$, with $e<0$ the charge of electron, while disorder is modelled by a short-range impurity potential $\hat{V}=\sum_{j}U\delta(\hat{\vec{r}}-\vec{R}_j)$, where $\vec{R}_j$ are randomly distributed impurity positions and $U$ characterises the scattering strength. Because the spin index $s=\uparrow,\downarrow$ is conserved, the problem separates into independent spin sectors, yielding two isotropic bands $\epsilon_{\vec{k}\eta}=\eta\mathcal{E}_{\vec{k}}$, 
where $\mathcal{E}_{k}=\sqrt{A^2k^2+(E^{-}_k)^2}$, where $E^{\pm}_k=M\pm Bk^2$. The corresponding eigenstates take the form $\vert \vec{k}s+ \rangle=\begin{bmatrix}
        s\cos\frac{\Theta_{k}}{2}e^{-si\theta_{\vec{k}}}&
        +\sin\frac{\Theta_{k}}{2}
    \end{bmatrix}^T$ and $\vert \vec{k}s- \rangle=\begin{bmatrix}
        s\sin\frac{\Theta_{k}}{2}e^{-si\theta_{\vec{k}}} &
        -\cos\frac{\Theta_{k}}{2}
    \end{bmatrix}^T$, 
with $\cos\Theta_{k}=E^{-}_k/\mathcal{E}_{k}$, $\sin\Theta_{k}=Ak/\mathcal{E}_{k}$, and $\theta_{\vec{k}}=\text{angle}\left(\frac{k_x+ik_y}{k}\right)$~\cite{zhang2026theoryti}.  Throughout this work, we consider a uniform electric field applied along the $x$ direction, $\vec{E}=E_{x}\vec{i}$.

%For equilibrium distribution function $\varrho^{\eta\eta'}_{\vec{k}s,0}=\delta_{\eta\eta'}f_{k\eta}$, we have vanishing collision integral, i.e.,   $\mathcal{ J}^{\bar{\eta}\eta}_{\vec{k}}=0$, where $f_{k\eta}=\frac{1}{e^{\beta\epsilon_{k\eta}}+1}$ is the Fermi-Dirac distribution function at inverse temperature $\beta=1/(k_BT)$.

%The previous theory of mesoscopic topological insulators, which focus on the topologically protected helical edge states, often omits the bulk contribution as it is an insulator~\footnote{The Drude contribution vanishes, since the density of states becomes zero at the Fermi energy.}. This electric field-drive coherence contributes to a quantized transverse charge Hall current even in the presence of a sizable gap in topologically nontrivial regime. This bulk contribution from quantum coherence is crucial, and difficult to separated from edge contribution in realistic experiments.

To incorporate the quantum decoherence -- the relaxation of the off-diagonal density matrix, we here go beyond the previous results of Ref.~\cite{culcer2017interband, culcer2022anomalous}, and assume the following ansatz for off-diagonal density matrix into the collision integral, which 
are described by two unknown complex parameters $\tau^{\bar{\eta}\eta}_{ks,\Vert}$ and $\tau^{\bar{\eta}\eta}_{ks,\perp}$, dubbed as ordinary and anomalous decoherence time~\cite{zhang2026theoryti}
\begin{align} \label{ansatz}
    \delta\varrho^{\bar{\eta}\eta}_{\vec{k}s}=\delta\varrho^{\bar{\eta}\eta}_{\vec{k}s,\Vert}+\delta\varrho^{\bar{\eta}\eta}_{\vec{k}s,\perp},
\end{align}
with
\begin{align} \label{fvfkvmkdf4}
\delta\varrho^{\bar{\eta}\eta}_{\vec{k}s,\Vert/\perp}=-\frac{e}{\hbar}\tau^{\bar{\eta}\eta}_{ks,\Vert/\perp}(f_{k\bar{\eta}}- f_{k\eta})\vec{\mathcal{R}}^{\bar{\eta}\eta}_{\vec{k}s}\cdot \vec{E}_{\Vert/\perp},
\end{align}
where $\vec{E}_{\Vert}=\vec{E}$,  $\vec{E}_{\perp}=\vec{E}\times \hat{z}$,  and $f_{k\eta}=1/(e^{\beta\epsilon_{k\eta}}+1)$ is the Fermi–Dirac distribution with inverse temperature $\beta=1/(k_BT)$ and energy spectrum $\epsilon_{\vec{k}\eta}$. The Berry connection is defined as $\vec{\mathcal{R}}^{\eta\eta'}_{\vec{k}s}=\langle \vec{k}s\eta\vert i\vec{\nabla}_{\vec{k}}\vert \vec{k}s\eta'\rangle$~\cite{xiao2010berry}.  Following the methodology of Refs.~\cite{zhang2026theory,zhang2026magnetoresistance,zhang2026theoryti}, we attain the spin-dependent normal and anomalous decoherence time  
\begin{align} \label{sgfbf1}
    \frac{\tau^{\bar{\eta}\eta}_{ks,\Vert}}{\hbar}=\frac{(\epsilon_{k\bar{\eta}}-\epsilon_{k\eta})-i\Gamma_{k}}{[(\epsilon_{k\bar{\eta}}-\epsilon_{k\eta})-i\Gamma_{k}]^2+(\Gamma^a_{k})^2},
\end{align}
\begin{align} \label{sgfbf2}
    \frac{\tau^{\bar{\eta}\eta}_{ks,\perp}}{\hbar}=\frac{s\eta\Gamma^a_{k}}{[(\epsilon_{k\bar{\eta}}-\epsilon_{k\eta})-i\Gamma_{k}]^2+(\Gamma^a_{k})^2}.
\end{align}
The ordinary-type symmetric scattering causes quantum decoherence, manifested as the decay of the off-diagonal elements of the density matrix and quantified by a normal decoherence rate $\Gamma_{k}=\frac{\hbar}{4\tau^{0}_{k}}\left[1+\frac{(E^{-}_k)^2}{\mathcal{E}^2_{k}}\right]$, where $1/\tau^0_{k}=(2\pi/\hbar)n_{\text{i}}\nu_{k}U^{2}$~\cite{zhang2026theoryti}. The density of state for each energy band is given by $\nu_{k}=\frac{\mathcal{E}_k}{2\pi\vert A^2-2BE^- _k\vert }$. More importantly, the novel-type antisymmetric scattering gives rise to an additional, anomalous contribution to decoherence, with rate $\Gamma_k^a=\frac{\hbar}{2\tau^{0}_{k}} \frac{E^{-}_k}{\mathcal{E}_k}$, generates an effective out-of-plane magnetic field that points in opposite directions for different spins~\cite{zhang2026theoryti}. In Berry-curvature-dominated systems, the electric field not only perturbs electrons near the Fermi energy away from the equilibrium Fermi–Dirac distribution~\cite{drude1900elektronentheorie}, but also induces quantum coherence throughout the entire Fermi sea~\cite{culcer2017interband,sekine2017quantum,atencia2022semiclassical}. The resulting spin-resolved current density is given by 
\begin{align} \label{fvakdfk}
    \vec{J}^{\alpha}_{s}=-\frac{e^2}{\hbar}\sum_{\vec{\vec{k}}\eta}i(\epsilon_{k\eta}-\epsilon_{k\bar{\eta}})\frac{\tau^{\bar{\eta}\eta}_{ks,\alpha}}{\hbar}(f_{k\bar{\eta}}- f_{k\eta})\vec{\mathcal{R}}^{\eta\bar{\eta}}_{\vec{k}}\vec{\mathcal{R}}^{\bar{\eta}\eta}_{\vec{k}s}\cdot \vec{E}_{\alpha},
\end{align}
with $\alpha=(\Vert,\perp)$. The velocity matrix elements are related to the Berry connection via $\hat{v}^{\eta\bar{\eta}}_{i,\vec{k}}=\frac{i(\epsilon_{\vec{k}\eta}-\epsilon_{\vec{k}\bar{\eta}})}{\hbar}\mathcal{R}^{\eta\bar{\eta}}_{i,\vec{k}}$.

\textit{Vanishing conduction from perfect coherence--} In a clean enough topological insulator, impurity scattering becomes ignorable. The off-diagonal component of density matrix, quantifying the quantum coherence describe the quantum coherence between conduction and valence bands, is  proportional to the Berry connection ($\vec{\mathcal{R}}^{\bar{\eta}\eta}_{\vec{k}s}$) and, in steady state, is given by $\delta\varrho^{\bar{\eta}\eta}_{\vec{k}s}=-\frac{f_{k\bar{\eta}}- f_{k\eta}}{\epsilon_{k\bar{\eta}}- \epsilon_{k\eta}}\vec{\mathcal{R}}^{\bar{\eta}\eta}_{\vec{k}s}\cdot e\vec{E}$~\cite{culcer2017interband,sekine2017quantum,atencia2022semiclassical}. Electric-field–induced interband coherence generates geometric (Berry curvature) velocity, which is intrinsically perpendicular to the external electric field~\cite{xiao2010berry}, so it cannot produce  longitudinal conductivity. Thus, this insulator in the absence of impurity is perfectly insulated at low enough temperature. A natural question is: when impurities destroy the perfect coherence, will the system generate a longitudinal response?

\begin{figure}[t!]
\begin{center}
\includegraphics[width=0.48\textwidth]{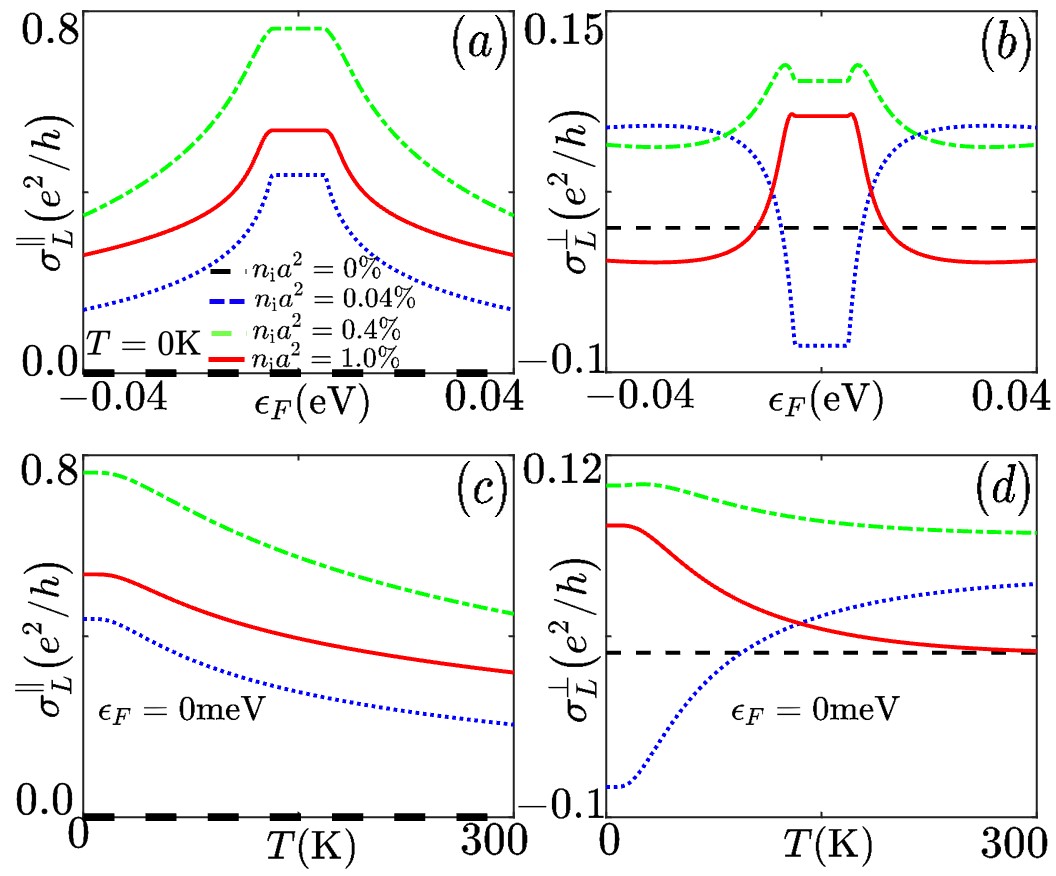} 
\end{center}
\caption{(a,b) Fermi energy ($\epsilon_F$) dependence of the longitudinal charge conductivities derived from (a) ordinary and (b) anomalous off-diagonal density matrix for different values of impurity density $n_{\text{i}}$. Panel (c) and (d) plot the corresponding temperature $T$ dependence.  Other parameters are from the experimental results Ref.~\cite{bernevig2006quantumexp}:  $M=5$meV, $A\pi/a=1.1$eV and  $B(\pi/a)^2=7.0$eV, $U/a^2=23$eV and $a=0.65$nm. }
\label{FIG7}
\end{figure}

\textit{Conduction from decoherence--} Here we  highlight that quantum decoherence can be a direct origin of electrical conduction in a bulk topological insulator with a sizable gap. This phenomenon can be understood by analogy with the double-slit experiment [Fig.~\ref{FIGS}(b)]: under perfectly coherent conditions, destructive interference initially supports a dark fringe (blue box), whereas the introduction of disturbances lifts this cancellation and gives rise to a bright fringe (red box). In this sense, quantum interference defines otherwise forbidden transport pathways, which become accessible once coherence is partially relaxed. In a topological insulator, this mechanism is especially striking. At zero temperature, the nonequilibrium population of states far from the Fermi energy is strongly suppressed; individual electrons or holes cannot be excited across the bulk gap to contribute to transport, resulting in a vanishing Drude response. By contrast, quantum coherence provides an alternative transport channel. When electron forms coherent superpositions of conduction- and valence-band states, its energy expectation remains centered at the Fermi energy, effectively bridging the huge bulk gap of topological insulators [Fig.~\ref{FIGS}(a)]. Once these coherent transport pathways become accessible, the decoherence process plays a dual role: while it disturbs quantum coherence, it converts the residual interband coherence into a finite and experimentally observable longitudinal response~\cite{zhang2026magnetoresistance}.

Quantitatively, the total longitudinal charge conductivity, derived from Eq.~\eqref{fvakdfk},  contain both ordinary and anomalous contributions, i.e.,  $\sigma^{}_{L}=\sigma^{\Vert}_{L}+\sigma^{\perp}_{L}$, where the longitudinal charge conductivity from $\delta\varrho^{\bar{\eta}\eta}_{\vec{k},\Vert}$ and  $\delta\varrho^{\bar{\eta}\eta}_{\vec{k},\perp}$, are given as follows
\begin{align} \label{f4jxdaroffxttTfyq}
    \sigma^{\Vert}_{L}&=-\frac{e^2}{h}\sum_{\eta}\eta\int^{\infty}_{0}kdkf_{k\eta} \text{Im}\left\{\frac{\tau^{+-}_{k,\Vert}}{\hbar}\right\} \frac{A^2}{\mathcal{E}_k}\left[1+ \frac{(E^{+}_k)^2}{\mathcal{E}^2_k}\right],
\end{align}
\begin{align} \label{f4jxdaroffxtptTfyq}
    \sigma^{\perp}_{L}&=-\frac{2e^2}{h}\sum_{\eta}\eta\int^{\infty}_{0}kdkf_{k\eta}\text{Re}\left\{\frac{\tau^{+-}_{k,\perp}}{\hbar}\right\}\frac{A^2E^{+}_k}{\mathcal{E}^2_k},
\end{align}
with  $\tau^{+-}_{k,\perp}=\tau^{+-}_{k\uparrow,\perp}$, where $f_{k\eta}=1/(e^{\beta\epsilon_{k\eta}}+1)$ is the Fermi–Dirac distribution with inverse temperature $\beta=1/(k_BT)$ and energy spectrum $\epsilon_{\vec{k}\eta}$. Figures~\ref{FIG7} (a) and (b) present the $\epsilon_{F}$ dependence of the conductivities  at zero temperature. We find $\delta\sigma^{\perp}_{L}$  can negative signs (blue curves), leading to a partial cancellation of the two decoherence-induced corrections. Both $\sigma^{\perp}_{L}$ and $\sigma^{\parallel}_{L}$ exhibit a plateau when $\epsilon_{F}$ lies within the bulk gap of the topological insulator. As $\epsilon_{F}$ is tuned within the gap, the available phase space for such interband coherence remains essentially unchanged, leading to the plateau in $\sigma^{\Vert/\perp}_{L}$. In contrast, once $\epsilon_{F}$ moves into the bands, the decoherence-induced conductivity shows a pronounced dependence on $\epsilon_{F}$ via its dependence on the Fermi–Dirac distribution [see Eqs.~\eqref{f4jxdaroffxttTfyq} and \eqref{f4jxdaroffxtptTfyq}]. Besides, temperature further modulates this behavior through the occupation factors. As shown in Figs.~\ref{FIG7}(c) and (d), the resulting temperature dependence is substantial, in stark contrast to the weak temperature variation of conventional Drude conductivity arising from static disorder. Importantly, in the high-temperature regime, one finds  $\sum_{\eta}\eta f_{k\eta}\simeq-\mathcal{E}_{k}/(2k_BT)$. The decoherence-induced contribution to the conductance is therefore inversely proportional to temperature, reflecting the fact that charge is carried by coherent superpositions of valence and conduction bands across the entire Fermi sea, rather than by Fermi-surface quasiparticles.  As shown in Eq.~\eqref{fvakdfk}, this unconventional scaling is expected to be \textit{universal} for coherent transport of Berry curvature-dominated system with impurity-scattering–induced coherence dissipation~\footnote{Decoherence is required, as a perfectly coherent state cannot support a longitudinal response.} and is reminiscent of strange-metal behaviour, most prominently observed in cuprate superconductors above their critical temperature~\cite{varma2020linear,cooper2009anomalous,jin2011link,licciardello2019electrical,bruin2013similarity,legros2019universal}.

\begin{figure}[t!]
\begin{center}
\includegraphics[width=0.48\textwidth]{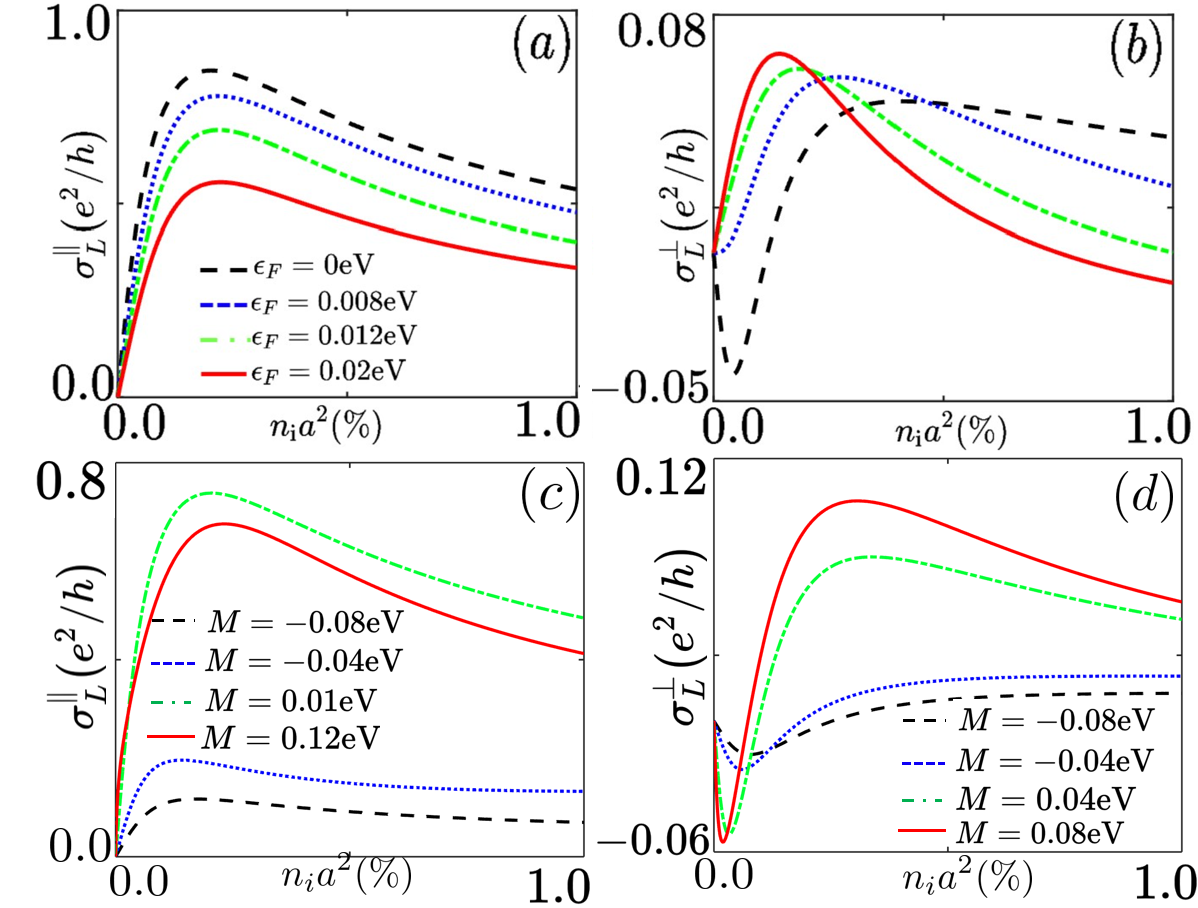} 
\end{center}
\caption{(a-b) The longitudinal  charge conductivities derived from
(a) ordinary and (b) anomalous off-diagonal density matrix as a function of impurity density $n_{\text{i}}$ for several values of Fermi energy ($\epsilon_F$). (c-d) The corresponding longitudinal  charge conductivities for several values of $M$ at 
$\epsilon_F=0$.
Other parameters are the same as Fig. \ref{FIG7}.}
\label{FIGM}
\end{figure}

Remarkably, an appreciable zero-temperature conductivity of order $\sigma_{L}\sim e^{2}/h$ persists when Fermi energy localized inside the gap of the topological insulator [see green curve in Figs.~\ref{FIG7}(a)], where the Drude contribution strictly vanishes. In this case,  Eqs.~\eqref{f4jxdaroffxttTfyq} and \eqref{f4jxdaroffxtptTfyq} reduce to 
\begin{align} \label{avfvrfre}
    \sigma^{\Vert}_{L}=\frac{e^2}{h}\int^{\infty}_{0}kdk \text{Im}\left\{\frac{\tau^{+-}_{k,\Vert}}{\hbar}\right\}\left[\frac{A^2}{\mathcal{E}_k}+ \frac{A^2(M+Bk^2)^2}{\mathcal{E}^3_k}\right],
\end{align}
\begin{align} \label{avfvrfvr}
    \sigma^{\perp}_{L}=\frac{2e^2}{h}\int^{\infty}_{0}kdk\text{Re}\left\{\frac{\tau^{+-}_{k,\perp}}{\hbar}\right\}\frac{A^2(M+Bk^2)}{\mathcal{E}^2_k}.
\end{align}
In the dilute impurity limit, one finds $\text{Im}(\tau^{+-}_{k,\Vert}/\hbar)\simeq \Gamma_k/4\mathcal{E}^2_k$ and $\text{Re}(\tau^{+-}_{k,\perp}/\hbar)\simeq -\Gamma^a_k/(4\mathcal{E}^2_k)$, both proportional to the impurity density $n_{\text{i}}$. Consequently, the decoherence-induced conductivity grows \textit{linearly} with impurity density $n_{\mathrm{i}}$ (i.e., $\sigma^{\parallel}_{L}\propto n_{\mathrm{i}}$ and $\sigma^{\perp}_{L}\propto n_{\mathrm{i}}$), as shown in Figs.~\ref{FIGM}(a) and (b). This scaling highlights that even a dilute concentration of impurities can generate a measurable longitudinal current via decoherence. The physical origin of this finite response is fundamentally distinct from conventional transport. Even in a gapped system with vanishing density of states at the Fermi energy, a conductivity of order $e^2/h$ survives because the effect is governed by the electrons of the  \emph{whole} Fermi sea rather than the only Fermi surface. Impurity-induced decoherence mixes conduction and valence band states of the same momentum, creating coherent superpositions that remain centered at the Fermi energy while effectively bridging the bulk gap [Fig.~\ref{FIGS} (a)]. As a result, the entire Fermi sea contributes to transport at zero temperature, as encoded in Eqs.~\eqref{avfvrfre} and \eqref{avfvrfvr}. Importantly, even when the Fermi energy lies within the bulk gap—particularly at low temperature—we find a finite conductivity in the topologically trivial regime (i.e., $MB<0$) [see blue and black curves in Fig.~\ref{FIG7}(c,d)], where helical edge states are absent. This unambiguously demonstrates that the residual conductivity originates from quantum decoherence, with edge contributions rigorously excluded. Our findings therefore provide a direct explanation for the experimentally observed nonzero longitudinal conductivity in the bulk gap at low temperatures~\cite{konig2007quantum,knez2010finite,shamim2021quantized}.

\begin{figure}[t!]
\begin{center}
\includegraphics[width=0.48\textwidth]{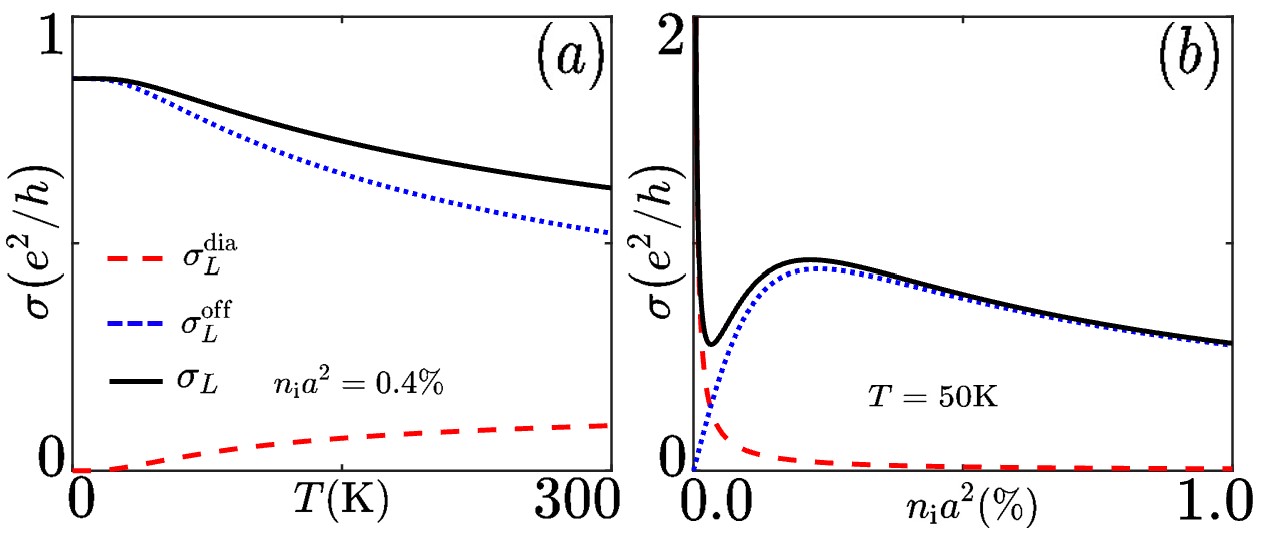} 
\end{center}
\caption{(a) $T$ dependence of $\sigma^{\text{dia}}_{L}$, $\sigma^{\text{off}
}_{L}=\sigma^{\Vert}_{L}+\sigma^{\perp}_{L}$, and $\sigma^{}_L=\sigma^{\text{dia}}_{L}+\sigma^{\text{off}}_{L}$. Panel (b) plots the corresponding $n_i$ dependence.  Other parameters are the same as Fig. \ref{FIG7}.}
\label{FIGC5}
\end{figure}

\textit{Comparison with MR from momentum relaxation--} Within Drude–Boltzmann transport theory~\cite{drude1900elektronentheorie,landau2013course,peierls1955quantum}, the Drude conductivity is governed by impurity scattering of the Fermi-surface quasiparticles. Disorder reduces quasiparticle lifetimes $\tau \propto 1/n_i $, yielding $\sigma_{\text{dia}} \propto \nu_F / n_i$~\cite{drude1900elektronentheorie}, with $\nu_F$ the density of states at the Fermi energy. Quantitatively, the Drude conductivity at finite temperature is given as follows
\begin{align} \label{fadvkfk}
    \sigma^{}_{\text{dia}}=-\frac{e^2}{4\pi}\sum_{\eta}\int^{\infty}_0 k dk v^2_{k\eta} \tau^{}_{k} f'_{k\eta},
\end{align}
with $v_{k\eta}=\eta k\frac{A^2-2B(M-Bk^2)}{\hbar \mathcal{E}_{k}}$ and $\tau^{-1}_{k}=\frac{2\pi}{\hbar}n_{\text{i}}\nu_{k}U^{2}\left(1-\frac{3A^2k^2}{4\mathcal{E}^2_k}\right)$.  Equation~\eqref{fadvkfk} yields $\sigma^{o}_{L}=0$ at $T=0$ reflecting the vanishing density of states at the Fermi energy. At elevated temperatures, thermal activation generates a finite carrier density in the topological insulator, leading to a small but finite Drude contribution, as illustrated by the red curve in Fig.~\ref{FIGC5}(a). In contrast, the decoherence-driven mechanism involves electrons throughout the entire Fermi sea, which participate collectively via interband quantum coherence. Figures~\ref{FIGC5}(a) and (b) show the dependence of the conductivity on $T$ and $n_{\text{i}}$, including the separate Drude and decoherence contributions as well as their sum. We find that, over a wide range of parameters, the total conductivity is predominantly governed by the decoherence-induced mechanism.

\textit{Conclusions—}We show that quantum decoherence generates a distinct longitudinal transport channel, yielding finite resistance even in the absence of carriers at the Fermi level. In the dilute limit, this contribution scales linearly with impurity density, in stark contrast to the Drude response.  This quantitative fingerprint enables experimental isolation of this effect.  The resulting quantum resistance establishes a direct and unambiguous signature of quantum decoherence, a fundamental limitation on the performance of nanoscale quantum devices. Our results establish quantum decoherence as a fundamental origin of longitudinal transport beyond the Drude paradigm, prompting a reexamination of the conventional classification of metals and insulators. Furthermore, the resistivity becomes linear in temperature at high temperatures. This unconventional scaling is expected to be universal for coherent transport and is a hallmark of the strange-metal regime observed in cuprate superconductors above the critical temperature near optimal doping. This  raises a natural question: is high-temperature superconductivity intrinsically linked to quantum coherence?

%More broadly, decoherence-induced quantum resistance offers a natural explanation for the ubiquitous observation of finite longitudinal conductivity even when the Fermi energy lies deep within the bulk gap, and extends to conventional semiconductors, where it may account for the pronounced conductivity enhancement induced by dilute impurities in systems such as silicon.

\textit{Acknowledgements--}
This work is supported by National Key R$\&$D Program of China (Grant Nos. 2020YFA0308800, 2021YFA1401500), the National Natural Science Foundation of China (Grant Nos. 12234003, 12321004, 12022416, 12475015, 11875108), and the National Council for Scientific and Technological Development (Grant No. 301595/2022-4).

\end{document}